\documentclass[sn-mathphys-num]{sn-jnl}

\usepackage{graphicx}%
\usepackage{multirow}%
\usepackage{amsmath,amssymb,amsfonts}%
\usepackage{amsthm}%
\usepackage{mathrsfs}%
\usepackage[title]{appendix}%
\usepackage{xcolor}%
\usepackage{textcomp}%
\usepackage{manyfoot}%
\usepackage{booktabs}%
\usepackage{listings}%

\begin{document}

\title[From directed percolation to patterned turbulence]{From directed percolation to patterned turbulence}


\author[1]{\fnm{Roger} \sur{Ayats}}\email{roger.ayatslopez@ist.ac.at}
\equalcont{These authors contributed equally to this work.}

\author[2]{\fnm{Lukasz} \sur{Klotz}}\email{lukasz.klotz@pw.edu.pl}
\equalcont{These authors contributed equally to this work.}

\author*[1]{\fnm{Björn} \sur{Hof}}\email{bhof@ist.ac.at}

\affil[1]{\orgname{Institute of Science and Technology Austria (ISTA)}, \orgaddress{\street{Am Campus 1}, \city{Klosterneuburg}, \postcode{3400}, \country{Austria}}}

\affil[2]{\orgdiv{Institute of Aeronautics and Applied Mechanics}, \orgname{Warsaw University of Technology}, \orgaddress{\street{Nowowiejska 24}, \city{Warsaw}, \postcode{00-665}, \country{Poland}}}

\abstract{
The transition to turbulence is characterized by an abrupt loss of order and predictability, featuring the intermittent proliferation and decay of localized turbulent structures \cite{avila2023transition,hof2023directed}. En-route to becoming fully turbulent, surprisingly order reappears when alternating laminar and turbulent regions arrange in regular stripe patterns \cite{tuckerman2020patterns}. This macroscopic organization is believed to arise top down from a classic pattern forming instability \cite{prigent2002large,kashyap2022linear} of turbulence, imprinting a wavelength onto the disordered flow field. We here demonstrate that patterns instead self-assemble with increasing velocity. Starting from the intermittent stripe regime, specifically from the corresponding directed percolation (DP) critical point, regular patterns are established within the scaling range of the DP transition. Likewise the patterns’ expansion rates are set by the DP critical exponents, attesting that all underlying processes are stochastic. This apparent contradiction between the inherent stochasticity and the displayed order is resolved by abandoning 
the common perception of laminar and turbulence as opposing states.
More generally our study exemplifies that macroscopic patterns can arise solely from local stochastic rules, in the absence of wavelength selection typically associated with pattern formation \cite{cross1993pattern,de2024pattern}.  

}


\maketitle

\begin{figure}                                                                
  \begin{center}
      \includegraphics[width=0.9\linewidth]{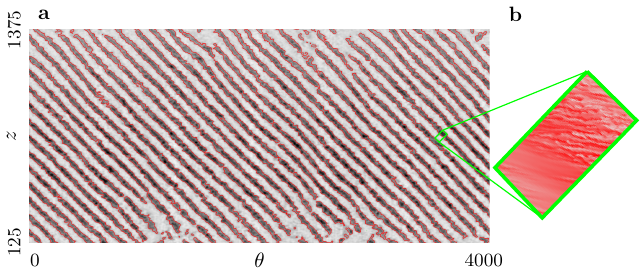}
  \end{center}  
  \caption{Laminar-turbulence patterns. Panel \textbf{a} shows the unfolded (i.e. azimuthal-axial plane) flow field  observed in a fluid layer sheared between two concentric cylinders for $Re=340$. Turbulent regions are marked red, whereas laminar regions appear in light grey. The flow structures are visualized using micron sized aluminium flakes. The spacing between stripes is approximately $70h$ where $h$ is half the thickness of the fluid layer, which corresponds to $2h=240 \mu m$, see \cite{klotz2022phase} for details. To highlight the interior turbulent dynamics, we show the flow field of a single stripe obtained from a direct numerical simulation for the same flow in \textbf{b}, computed in the small aspect ratio box shown.}  
\label{fig:general_visu}
\end{figure}

The co-existence of laminar and turbulent regions, commonly referred to as spatio-temporal intermittency, is intrinsic to the onset of turbulence and can be readily observed in a wide range of situations including flows through pipes \cite{reynolds1883iii}, boundary layers \cite{emmons1951laminar}, channels \cite{carlson1982flow} and fluid layers sheared between parallel walls \cite{prigent2002large,klotz_couette-poiseuille_2017,klotz_experimental_2021} or concentric cylinders \cite{coles1965transition,van1966exploratory,meseguermellibovsky2009}, respectively Couette and Taylor-Couette flows. The latter cases feature stripe patterns which are strikingly regular (e.g., Fig.~\ref{fig:general_visu}), contrasting the highly chaotic motion in the stripes' interior. This unlikely manifestation of order in turbulence, inspired Feynman \cite{Feynman} to highlight this phenomenon 
as an example illustrating humans’ inability to comprehend the qualitative content of differential equations. 
Indeed, classic stability analysis of the governing equations only confirms the linear stability of the laminar base flow but provides no insight into the onset of turbulence, let alone stripe patterns. 

On the other hand, taking the point of view of turbulence and by determining its lower Reynolds number limit, recent studies established that this reverse transition to laminar flow is continuous and falls into the directed percolation universality class \cite{lemoult2016directed,shih2016ecological,barkley2016theoretical,chantry2017universal,hiruta2020subcritical,takeda2020intermittency,klotz2022phase,lemoult2024directed}. Despite this progress, the DP framework falls short of explaining the whole range of the transition and is believed to 
only cover the lower end, from the sparse stripe / puff regime, down to the laminar absorbing state \cite{manneville2017laminar}. 
Within this regime, turbulence can only exist locally as it relies on energy input from the surrounding laminar flow \cite{van2009flow}. More precisely, laminar fluid of high kinetic energy continuously collides with the slower moving turbulent patch and, in doing so, sustains its interior eddying motions. Conversely the fluid exiting a turbulent patch has an energetically depleted velocity profile (plug-like in channels and pipes). With distance from the laminar-turbulent interface the velocity profile gradually (i.e., viscously) regains its more energetic laminar shape. This development length, sets a minimum spacing \cite{hof2010eliminating} between puffs in pipe flow, a length scale which matches the spacing between stripes \cite{samanta2011experimental}. Essentially it is this energetic depletion which prohibits streamwise extended turbulent structures, i.e., structures that exceed the length of a puff or the width of a stripe, to arise in the DP regime. 

Directed percolation on the other hand, being stochastic in nature, is believed to be of no relevance in the dense stripe regime. The ordered stripe patterns (e.g. Fig.~\ref{fig:general_visu}) observed are considered a separate phenomenon \cite{manneville2017laminar}, and  
are thought to arise top down, from an instability of fully turbulent flow. 
This instability of turbulence has been suggested to be either of Ginzburg-Landau 
\cite{prigent2002large,prigent2003long,berghout2020direct} or Turing type \cite{manneville2012turbulent,kashyap2022linear,kashyap2024laminar}. Recently \cite{benavides2023model} a model exhibiting an instability to stripe patterns from turbulence has been derived from a truncation of Waleffe flow. 

We will show in the following that, in contrast to this established top down view, dense stripe patterns already form within the scaling range of directed percolation, and hence a regime that is entirely governed by stochastic rules. 
A simple change of perspective reveals that what is commonly perceived and interpreted as a globally ordered pattern, is the order parameter one limit of the active fluctuating phase. Key to resolving the conflict between order and stochasticity is the realisation that the active phase of the DP transition is not turbulence, but a composite structure consisting of intense turbulence and the recovering, quasi laminar fluid driving it. Given that individual structures take the form of stripes, a dense tiling of this single phase naturally results in the stripe patterns observed. 

Direct numerical simulations of Couette flow are carried out in a long slender domain tilted by $24^{\circ}$ with respect to the streamwise direction, following previous works (e.g, \cite{tuckerman2011patterns,shi2013scale,lemoult2016directed}). The dimensions of the periodic computational domain range from ${(L_x,L_y,L_z)=(10,2,40)h}$ to ${(L_x,L_y,L_z)=(10,2,960)h}$, where $2h$ corresponds to the gap between the parallel walls, and $L_x$, $L_y$ and $L_z$ are the dimensions in the directions parallel, wall-normal and perpendicular to the stripe, respectively. The Navier–Stokes and continuity equations for an
incompressible Newtonian fluid are numerically solved by using a pseudo-spectral solver for planar geometries adapted from Openpipeflow \cite{willis2017openpipeflow}.  The spatial resolutions range from ${(N_x,N_y,N_z)=(24,27,128)}$ to ${(N_x,N_y,N_z)=(24,27,3072)}$, $N_x$ and $N_z$ being the number of Fourier modes in the streamwise and spanwise directions, and $N_y$ the number of grid points in the wall-normal direction.
\begin{figure}[h]                                                                 
  \begin{center}
  \includegraphics[width=0.7\linewidth]{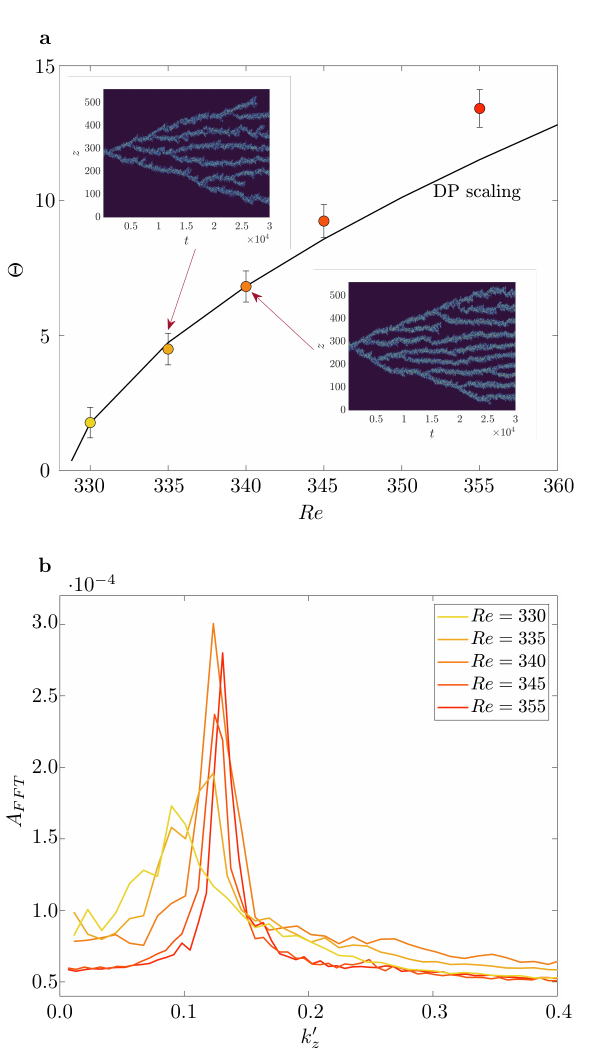}\\
  \end{center}  
  \caption{Emerging stripe pattern. Panel \textbf{a} shows the opening angle of the ensemble averaged spreading cones as a function of $Re$. Two such spreading cones are shown in the insets. The scaling predicted from directed percolation is given by the black line. The spatial Fourier transform as a function of the wavenumber $k'_z$ shown in panel \textbf{b} confirms the emergence of a wavelength within the DP scaling regime.  The analysis is carried out using the wall normal kinetic energy, $A(x,z)={0.5\,u_y(x,z)}^2$, $u_y(x,z)$ being the the wall-normal velocity averaged along half of the gap width ${y=[-1,0]}$. Next we obtain  ${A_{FFT}(z)=\langle A(x,z)\rangle_{xt}}$, where ${\langle \cdot{} \rangle_{xt}}$ denotes spatial average along the direction $x$ (parallel to the stripe) and time ensemble average.}
\label{fig:Afft}
\end{figure}
As shown in earlier simulations of the identical geometry, the directed percolation phase transition to sustained stripes in this case occurs at $Re_c\approx328.7$  \cite{lemoult2016directed}. Starting from this point, we conducted simulations at ${Re=[330,335,340,345,350]}$. In each case the flow is initiated with a single stripe and continued until stripes (via stripe splitting) have spread throughout most of the domain. Ensemble averages of such single seed experiments result in a spreading cone with a well defined opening angle, which increases with distance from the critical point. Within the directed percolation scaling range, this angle is proportional to the ratio of the spatial and the temporal correlation exponents $\nu_\perp$ and $\nu_\parallel$, respectively \cite{hinrichsen2000non}. 
For each $Re$ listed above we performed ten simulations starting from different single stripe conditions, and the measured ensemble averaged opening angle is shown as a function of $Re$ in Fig.~\ref{fig:Afft}a (colored bullets). For comparison, the directed percolation prediction is given by the black line. The data follows the predicted angle up to $Re=340$ and slowly deviates starting from $Re=345$. Visual inspection reveals that surprisingly the interior of the cones is not entirely irregular and disordered as would be expected for directed percolation. Already from $Re\geq335$ a characteristic length scale separating adjacent stripes is perceived, and this spacing matches what previous studies including simulations of the same tilted geometry at identical parameter values \cite{tuckerman2011patterns} interpreted as a wavelength originating from an instability of fully turbulent flow. This standard pattern formation interpretation\cite{cross1993pattern,prigent2002large} appears to be supported by a simple spatial Fourier analysis shown in Fig.~\ref{fig:Afft}b. The fast Fourier transform of the local wall-normal kinetic energy (see caption Fig.~\ref{fig:Afft} for details), reveals the emergence of a singe peak from $Re\geq335$ and this peak saturates at a high level from $Re=340$ when, again in agreement with previous studies of the same flow \cite{tuckerman2011patterns}, stripe patterns dominate the flow field. However, at the same time as shown in Fig.~\ref{fig:Afft}b, $Re=340$ is within the scaling range of the DP transition. Therefore, a globally selected wavelength is incompatible with the stochasticity as well as with the strict locality of the processes underlying DP.

While above simulations have been conducted in long but slender tilted domains and stripe patterns are therefore quasi one dimensional (1D), we next switch to large aspect ratio experiments where patterns are naturally two dimensional. Specifically, experiments are conducted in circular Couette flow, i.e., for a fluid layer between two concentric cylinders with an azimuthal aspect ratio of $L_{\theta}=C/h=3927$ and an axial aspect ratio $L_z=H/h=1500$, where C and H are the azimuthal circumference and cylinder height and $2h$ is the gap between the two cylinders. Further details of the experimental procedures can be found in Klotz et al. \cite{klotz2022phase}. As demonstrated in this latter study, the transition falls into the universality class of directed percolation in 2+1 dimensions and the critical point corresponds to $Re=330$. A typical visualization image of the flow field in the azimuthal-axial plane, $3\%$ above critical ($Re=340$) is shown in Fig.~\ref{fig:general_visu}.

\begin{figure}[h]                                                                 
  \begin{center}
   \includegraphics[width=0.7\linewidth]{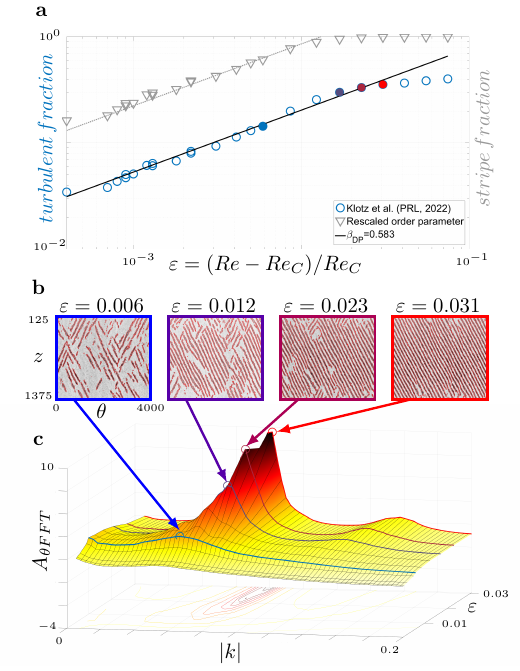}
  \end{center}  
  \caption{Turbulent fraction as a function of $\varepsilon$ (panel \textbf{a}); instantaneous patterns for $\varepsilon \in (0.006,0.031)$ (panel \textbf{b}); amplitude of spatial Fourier transform $A_{FFT \theta}$ as a function wavenumber perpendicular to the pattern $|k|$ and $\varepsilon$ (panel \textbf{c}). The colors of the filled symbols in panel \textbf{a,} correspond to the $\varepsilon$ values of the frames selected in panel \textbf{b}, and to the highlighted (same color coding) spectra in panel \textbf{c}} 
\label{fig:AfftExp}
\end{figure}

\begin{figure}[h]                                                      
  \begin{center}
            \includegraphics[width=0.7\linewidth]{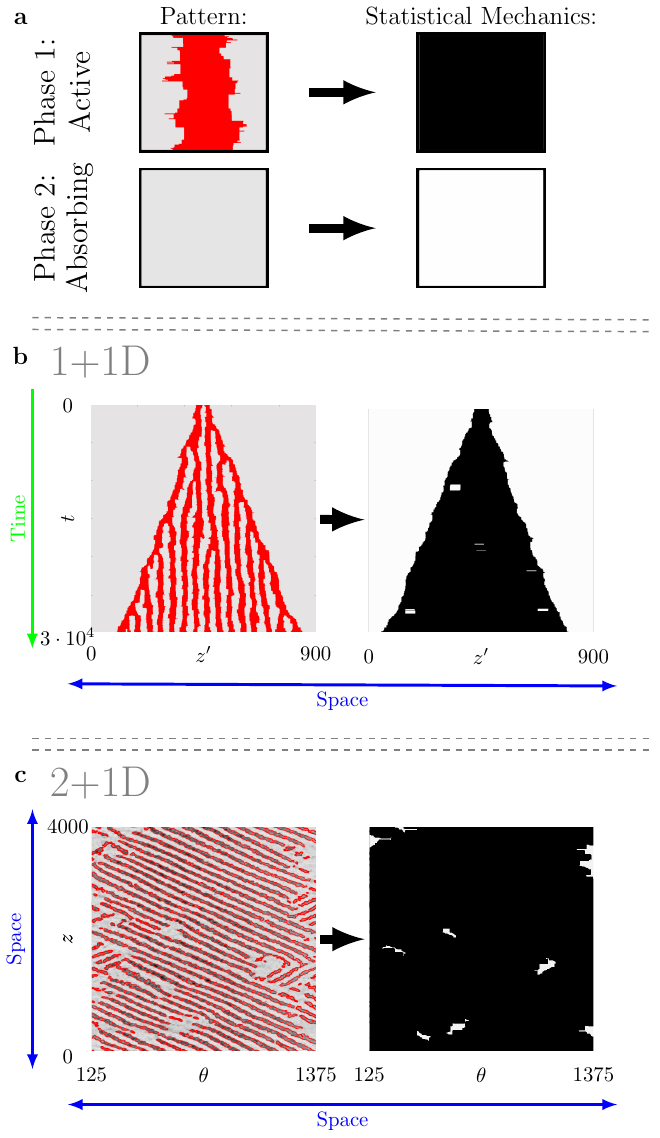} 
  \end{center}  
  \caption{Mapping from flow fields to phases. Panel \textbf{a} illustrates the two phases of the directed percolation transition. Notably the active phase is a hybrid consisting of the turbulent core, driven by the adjacent laminar fluid. For a stripe angle of $24^{\circ}$ the minimum size of an active site corresponds to $40h$. Absorbing sites must in principle be excitable to the active phase via neighbour interactions and this is only possible if the laminar gap has a width of $40h$. Panel \textbf{b} shows the mapping for the spreading cone for the simulations of stripes in the one dimensionally extended domain. Panel \textbf{c} shows the equivalent mapping of the flow field at $\epsilon=0.023$ to the phases. As this mapping reveals, in both cases the perceived stripe patterns correspond to densely packed active sites and hence in the statistical mechanics context to a virtually fully occupied domain.}
\label{fig:cones}
\end{figure}

We analyse flow fields in the immediate vicinity of the critical point, where the turbulent fraction scales with the universal critical exponent of $\beta\approx0.58$. As shown in Fig.~\ref{fig:AfftExp}, dense stripe patterns and a distinct peak of the respective Fourier transform appear within the DP scaling range. Here, in accordance with the DP conjecture \cite{hinrichsen2000non}, stripes correspond to the fluctuating active phase. Fluctuating in this context refers to the stochastic generation and disappearance of sites, i.e., the decay and splitting of stripes. These memoryless stochastic rules and the strictly local interactions, just like in the 1D case, are incompatible with a pattern forming instability and a global wavelength.  
At the same time, the intrinsically stochastic processes that dominate this flow regime appear at odds with the ordered patterns that emerge. For directed percolation order can only be resumed in the limit of the order parameter being one, i.e., for the trivial case of a fully occupied domain.

Key to resolving this apparent contradiction between percolation and patterns lies in the naturally assumed correspondence between turbulence and active DP sites. If correct, two such turbulent sites, just like two active DP sites, would have to be able to exist in direct proximity, resulting in an active region twice as large. 
However, and as discussed above, the turbulent core of a stripe (/puff) has a well defined mean size and neighbouring stripes are separated by laminarescent fluid. As in the case of adjacent puffs, this separation zone has to be sufficiently large to allow the local velocity profile to recover towards a more energetic shape, in order to drive turbulence. Conversely, if two stripes(see SImovie) are separated by less than the required recovery distance, one stripe decays. Equally a profile distortion (not shown) that inhibits the recovery of the fluid between stripes causes the stripes' immediate decay. 

Directed percolation describes the transition from an active phase into an absorbing phase, where each phase must exist independent of its neighbours (in space). As we just saw, this requirement is not met by the turbulent core of a stripe/puff. Thus, the smallest active unit that is self-sustained is a stripe, including the recovering fluid around it.  
An active site is therefore a composite structure which has a specific width dictated by the profile recovery, shown in Fig.~\ref{fig:cones}a (left). For a stripe angle of $24^{\circ}$ the minimum domain width for such a self sustained stripe corresponds to $40h$ (see also \cite{tuckerman2020patterns}). Equally, an absorbing site has to be of the appropriate size so that in principle it can be occupied by a sustained stripe  in subsequent time steps, resulting in the identical size requirement of $40h$.

With this in mind, we convert the stripe images to the actual phases of the transition by marking active sites in black and absorbing sites in white (Fig.~\ref{fig:cones}a). The resulting spreading cones for the quasi 1D flow are shown in Fig.~\ref{fig:cones}b. 
This simple conversion from the dynamical states (laminar, turbulent) to the phases of the transition clarifies that the perceived pattern corresponds to a fully occupied region. It is physically impossible to fit additional turbulence into the perceived gaps between densely spaced stripes. The identical argument holds for the two dimensional patterns in experiments, as shown in Fig.~\ref{fig:cones}c. Equally here the active phase occupies almost the entire domain. Wider gaps (i.e. empty sites) that occasionally arise between stripes are quickly closed by the generation of new stripes via splittings. Returning to Fig.~\ref{fig:AfftExp}a and redefining the order parameter as the stripe fraction (instead of the turbulent fraction), shows that the corrected values are larger than previously expected. The order parameter one limit (see light grey symbols in Fig.~\ref{fig:AfftExp}a) of this transition then corresponds to a dense stripe pattern. 

In conclusion, the onset of turbulence is by default considered a transition between laminar and turbulent motion. However, counter-intuitively these two dynamical states do not correspond to the two phases of the underlying directed percolation transition. Turbulence is not stable in isolation and therefore does not qualify as a phase. Instead the active phase is a composite of laminar and turbulent motion. 
The order parameter one limit, i.e. a dense tiling of this active phase, naturally corresponds to an ordered sequence of stripes. While our finding does not rule out a pattern forming instability at larger $Re$, the global wavelength associated with such a potential instability is incompatible with directed percolation and cannot prevail in the DP scaling regime and consequently would be irrelevant for the parameters of the present study. 

More generally our findings demonstrate that the patterns observed in the Re regime studied, neither require classical linear wavelength selection \cite{cross1993pattern}, nor recent extensions to nonlinear processes \cite{de2024pattern} but that instead, patterns can arise in the complete absence of wavelength selection, and self-assemble even from fully stochastic local interactions.

\bmhead{Acknowledgements}
We thank Gökhan Yaln{\i}z for contributing with initial simulations and fruitful discussions. 
R.A. was supported by the European Union’s Horizon 2020 research and innovation program under the Marie Sk\l{}odowska-Curie Grant Agreement No. 101034413, and by the Austrian Science Fund (FWF) 10.55776/ESP1481224. L.K. was supported by the European Union’s Horizon 2020 research and innovation program under the Marie Sk\l{}odowska-Curie Grant Agreement No. 754411, and by the National Science Center (Poland) within OPUS-21 project (2021/41/B/ST8/03142).



\bibliography{ref}

\end{document}